\documentclass[aps]{revtex4}
\usepackage{epsfig}
\usepackage{graphicx}
\begin{document}
\title{Aspects of short range correlations in a relativistic model}
\author{P.K. Panda}
\author{D.P. Menezes}
\affiliation{Depto de F\'isica-CFM, Universidade Federal de Santa
Catarina, CP. 476, 88040-900 Florian\'opolis-SC, Brazil}
\author{C. Provid\^encia}
\author{J. Da Provid\^encia}
\affiliation{Centro de F\'isica Te\'orica - Dep. de F\'isica,
Universidade de Coimbra, P-3004 - 516 Coimbra, Portugal}
\begin{abstract}

In the present work short range correlations are introduced for the first time
in a relativistic approach to the equation of state of the
infinite nuclear matter in the framework of the Hartree-Fock
approximation using an effective Hamiltonian derived from the
$\sigma-\omega$ Walecka model. The unitary correlation method is used
to introduce short range correlations. The effect of the correlations in the
ground state properties of the nuclear matter is discussed.
\end{abstract}
\maketitle
Nuclear physics is an effective theory of the nucleus regarded as
a system of nucleons. In this theory the essential degrees of
freedom are the center of mass coordinates of the nucleons as
well as their spins and isospins, the interactions being
expressed as nucleon-nucleon forces. It is usually accepted,
however, that the nucleon-nucleon interaction becomes strongly
repulsive at short distances in the relative coordinate of two
nuclear particles. It is seen that the phase shifts 
for $^1S_0$ and $^3S_1$ are positive at low and becomes negative 
at higher energies \cite{macgregor}. This indicates a repulsive core at short
distances and attraction at long distances. 
There are attempts to derive the
nucleon-nucleon force using chiral perturbation theory
\cite{chpt}. In this approach,  one- and two-pion exchange contributions
are taken into account up to the third chiral order. However, in order to
reproduce the nucleon scattering data and the D-wave phase shifts by this
method, an {\it ad hoc}  contact interaction (which represents the short
range force) must also be included. The realistic nucleon-nucleon
forces are basically phenomenological. The Bonn interactions
\cite{bonn} are based on meson exchange which is treated in a
relativistic non-local manner. The Argonne interactions
\cite{argonne}, on the other hand, describe the pion exchange in a
local approximation, the short and medium range nuclear interactions
being controlled by phenomenological parameters.

Nonrelativistic calculations based on realistic NN potentials predict
equilibrium points that are not able to describe simultaneously the correct
binding energy and saturation density; either the saturation
density is correct but the  binding energy is too small, or the correct binding
energy is obtained at a too high density. This behaviour is normally referred
to ``Coester line'' \cite{Coester}.
This problem is generally circunvented through the introduction of a
three-body repulsive force \cite{3-body}, or density-dependent repulsive
mechanisms. A mechanism of this kind is already present in relativistic models.
Due to Lorentz covariance and self-consistency, as the nuclear density
increases, the nucleon effective mass decreases. As a result there is a
reduction of the  attractive force  and a net increase of the repulsive
force. In fact, the relativistic mean field (RMF) theory formulated by Teller
and the others \cite{teller} and by Walecka \cite{walecka} is 
successful in
describing both infinite nuclear matter and finite nuclei.
Moreover, the relativistic Hartree-Fock (HF) approximation 
\cite{horowitz,bouyssy}
has also been used for the description of finite nuclei and infinite nuclear
matter.

In this paper, we have focused on the treatment of short range correlations
in dense nuclear matter for the first time using
relativistic equations of motion. Although there are several procedures which
may be used to introduce short range correlations into the model wave
function, we have preferred to work with the unitary operator method as
proposed by Villars \cite {villars}. There are several advantages in using an
unitary model operator. In particular, one automatically guarantees that the
correlated state is normalized. Further, we shall see that for
calculation of correlation corrections to the matrix elements of one- and 
two-body operators is a simple task. The general idea exists for a long
time \cite{jp,suzuki,feldmeier} but has not been pursued for the
relativistic case. This has been so because in non-relativistic
models the interaction arises from the interplay between a long
range attraction and a very strong short range repulsion so that,
indeed, it is indispensable to take short range correlations into
account.  In relativistic mean field models, 
the parameters are phenomenological, fitted to the
saturation properties of nuclear matter.  Short
range correlation effects are clearly included in the model parameters.
We expect that the values of these parameters are more 'fundamental', or
less artificial, when  correlations are not neglected.

We start with the effective Hamiltonian as
\begin{eqnarray}
H&=&\int \psi^\dagger_\alpha (\vec x)(-i\vec\alpha\cdot\vec\nabla
+\beta M)_{\alpha\beta}\psi_\beta(\vec x)~d\vec x\nonumber\\
&+&\frac{1}{2}\int\psi_\alpha^\dagger (\vec x)\psi_\gamma^\dagger (\vec y)
V_{\alpha\beta,\gamma\delta}(|\vec x -\vec y|)
\psi_\delta(\vec y)\psi_\beta(\vec x)~d\vec x~d\vec y
\end{eqnarray}
with
\begin{equation}
V_{\alpha\beta,\gamma\delta}(r)=
(\beta)_{\alpha\beta}(\beta)_{\gamma\delta}V_\sigma(r)
+\left(\delta_{\alpha\beta}\delta_{\gamma\delta}
-\vec\alpha_{\alpha\beta}\cdot\vec\alpha_{\gamma\delta}\right)V_{\omega}(r)
\end{equation}
where
\begin{equation}
V_\sigma(r)=-\frac{g_\sigma^2}{4 \pi}\frac{e^{-m_\sigma
r}}{r},\quad\quad V_{\omega}(r)=\frac{g_\omega^2}{4
\pi}\frac{e^{-m_\omega r}}{r},
\end{equation}
and $\vec\alpha$ are the  Dirac-matrices.
In the above, $\psi$ is the nucleon field interacting through the scalar
and vector potentials.
The equal time quantization condition for the nucleons is given by
\begin{equation}
[\psi _{\alpha}(\vec x,t),\psi _\beta (\vec y,t)^{\dagger}]_{+}
=\delta _{\alpha \beta}\delta(\vec x -\vec y),
\end{equation}
where $\alpha$ and $\beta$ refer to the spin indices. We now also have the
field expansion for the nucleons
$\psi$ at time t=0  given as \cite{mishra}
\begin{equation}
\psi(\vec x)=\frac {1}{\sqrt{V}}\sum_{r,k} \left[U_r(\vec k)c_{r,\vec k}
+V_r(-\vec k)\tilde c_{r,-\vec k}^\dagger\right] e^{i\vec k\cdot \vec x} ,
\end{equation}
where $U_r$ and $V_r$ are given by
\begin{equation}
U_r(\vec k)=\left( \begin{array}{c}\cos\frac{\chi(\vec k)}{2}
\\ \vec \sigma \cdot\hat k\sin\frac{\chi(\vec k)}{2}
\end{array}\right)u_r~ ;~~ V_r(-\vec k)=\left(
\begin{array}{c}-\vec \sigma \cdot\hat k\sin\frac{\chi(\vec k)}{2}
\\ \cos\frac{\chi(\vec k)}{2}\end{array}\right)v_r~.
\end{equation}
For free spinor fields, we have $\cos\chi(\vec k)=M/\epsilon(\vec
k)$, $\sin\chi(\vec k)=|\vec k|/\epsilon(\vec k)$ with
$\epsilon(\vec k)=\sqrt{\vec k^2 + M^2}$. However, we will deal
with interacting fields so that we take the ansatz $\cos\chi(\vec
k)=M^*(\vec k)/\epsilon^*(\vec k)$, $\sin\chi(\vec k)=|\vec
k^*|/\epsilon^*(\vec k)$, with $\epsilon^*(\vec k)=\sqrt{\vec
{k^*}^2 + {M^*}^2(\vec k)}$, where $\vec k^*$ and $M^*(\vec k)$
are the effective momentum and effective mass respectively. The
equal time anti-commutation conditions are $ [c_{r,\vec
k},c_{s,\vec k'}^\dagger]_{+}~=~ \delta _{rs}\delta_{\vec k,\vec
k'}~=~ [\tilde c_{r,\vec k},\tilde c_{s,\vec k'}^\dagger]_{+}~. $
The vacuum $\mid 0\rangle$ is defined through $c_{r,\vec k}\mid
0\rangle=\tilde c_{r,\vec k}^\dagger\mid 0 \rangle=0$;
one-particle states are written $|\vec k,r\rangle=c_{r,\vec
k}^\dagger\mid 0\rangle$; two-particle and three-particle
uncorrelated states are written, respectively as $|\vec k,r;\vec
k',r'\rangle= c_{r,\vec k}^\dagger ~c_{r',\vec k'}^\dagger\mid
0\rangle$, and $|\vec k,r;\vec k',r'; \vec k'',r''\rangle=
c_{r,\vec k}^\dagger ~c_{r',\vec k'}^\dagger ~c_{r'',\vec
k''}^\dagger \mid 0\rangle,$ and so on.

\begin{figure}
\includegraphics[width=12.cm,angle=0]{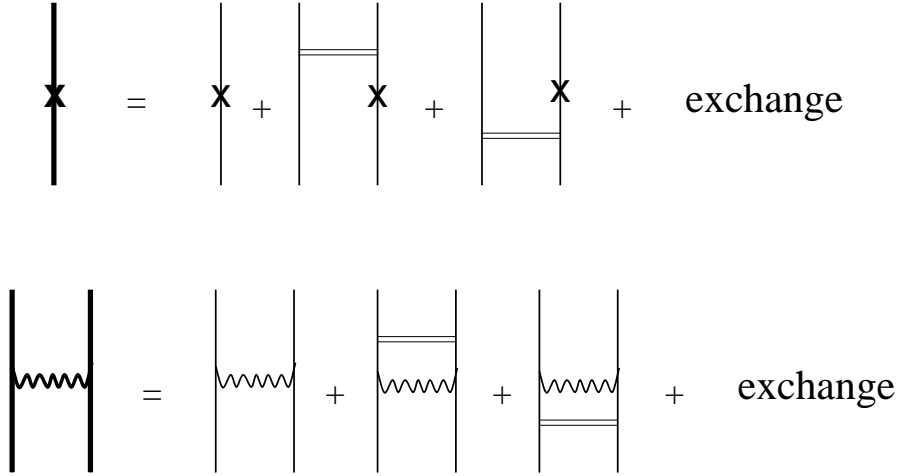}
\caption{The kinetic and the potential contribution to the two-body cluster 
diagrams}
\label{diagram}
\end{figure}

We now introduce the short range correlation
through an unitary operator method. The correlated wave function is
\cite{jp1} $|\Psi\rangle=e^{iS}|\Phi\rangle$ where $|\Phi\rangle$
is a Slater determinant and $S$ is, in general, a n-body
Hermitian operator, splitting into a 2-body part, a 3-body part,
etc.. Consider now the expectation value of $H$,
\begin{equation}
E=\frac{\langle \Psi| H|\Psi\rangle}{\langle \Psi |\Psi\rangle}=
\frac{\langle \Phi| e^{-iS}~H~e^{iS}|\Phi\rangle}{\langle \Phi|
\Phi\rangle}.
\end{equation}
At this point we restrict ourselves to the two-body correlation diagrams shown
in figure \ref{diagram}.
Let us denote the  two-body  correlated wave function by
\begin{equation}
|\overline{\vec
k,r;\vec k',r'}\rangle=e^{iS}|{\vec k,r;\vec k',r'}\rangle\approx
f_{12}|{\vec k,r;\vec k',r'}\rangle
\end{equation}
where $f_{12}$ is the short range correlation factor, the
so-called Jastrow factor \cite{jastrow}. For simplicity, we have
written
$f_{12}=f(\vec r_{12})$, $\vec r_{12}=\vec r_1-\vec r_2$.
We take the assumption that
$f(r)=1-(\alpha +\beta r)~e^{-\gamma r}$
where $\alpha$, $\beta$ and $\gamma$ are parameters.

At this point some remarks on the choice of the Jastrow factor
are appropriate. The important effect of the short range
correlations is the replacement, in the expression for the
ground-state energy, of the interaction matrix element
$\langle{\vec k,r;\vec k',r'}|V_{12}|{\vec k,r;\vec k',r'}\rangle$
by $\langle\overline{\vec k,r;\vec
k',r'}|V_{12}+t_1+t_2|\overline{\vec k,r;\vec
k',r'}\rangle-\langle{\vec k,r;\vec k',r'}|t_1+t_2|{\vec k,r;\vec
k',r'}\rangle$, where $t_i$ is the kinetic energy operator of
particle $i$. As argued by Moszkowski \cite{mosz} and Bethe \cite{bethe},
it is expected that the true ground-state wave
function of the nucleus containing correlations coincide
with the independent particle, or Hartree-Fock wave function, for interparticle
distances $r\geq r_{heal}$, where $r_{heal}\approx 1$ fm is the
so-called ``healing distance". This behavior is a consequence
of the constraints imposed by the Pauli Principle. Moreover,
although in general the correlation factor $f(r)$ may depend on
the isospin and spin quantum numbers of the two-body channel, we
assume, for simplicity, that it is a plain, state independent
Jastrow factor. A natural consequence of having the
correlations introduced by an unitary operator is the occurrence of
a normalization constraint on $f(r)$,
\begin{equation}
\int~(f^2(r)-1)~d^3r=0.\label{c0}
\end{equation}

The correlated ground state energy becomes
\begin{eqnarray}
{\cal E}&=&\frac{\nu}{\pi^2}\int_0^{k_F} k^2 ~dk~
\left[|k|\sin\chi(k)~+~M\cos\chi(k)\right]
~+~\frac{\tilde F_\sigma(0)}{2}\rho_s^2 +\frac{\tilde
F_{\omega}(0)}{2}\rho_B^2\nonumber\\
&+& C_0 \rho_B \frac{\nu}{\pi^2}\int_0^{k_F} k^2 ~dk~ \left[
|k| \sin\chi(k)~+~ M\cos\chi(k) \right]\nonumber\\
&-&\frac{4}{(2\pi)^4} \int_0^{k_f} k^2~ dk~ {k'}^2~ d
k'~ \left\{\Big[ |k| \sin\chi(k)+ 2~M\cos\chi(k)\Big] I(k,k')
+ |k|~\sin\chi(k')~J(k,k') \right\}\nonumber\\
&+&\frac{1}{(2\pi)^4}\int_0^{k_f} k~ dk~ k'~ dk'
\left[\sum_{i=\sigma,\omega} A_i (k,k')
+\cos\chi(k)\cos\chi(k') \sum_{i=\sigma,\omega}B_i (k,k') 
+ \sin\chi(k)\sin\chi(k')
\sum_{i=\sigma,\omega} C_i(k,k')\right]\nonumber\\
\end{eqnarray}
where, $C_0=\int~(f^2(r)-1)~d^3r$ so that, according to
(\ref{c0}), $C_0=0,$ and $A_i$, $B_i$, $C_i$ , $I$ and $J$ are
exchange integrals. In the above equation for the energy density, the
first term results from the kinetic contribution, the second and third 
terms come respectively from the $\sigma$ and $\omega$ direct contributions
from the potential energy with correlations, the fourth and
fifth ones from the direct and exchange correlation contribution
from the kinetic energy, and the last one from the $\sigma +
\omega$ exchange contributions from the potential energy with correlations. The
angular integrals are given by 
$A_i(k,k')=B_i(k,k')=2\pi ~g_i^2/4\pi\int_0^\pi d \cos \theta
~\tilde F_i(k,k',\cos \theta),$
$C_i(k,k')=2\pi ~g_i^2/4\pi\int_0^\pi \cos \theta ~d \cos \theta
~\tilde F_i(k,k',\cos \theta),$ 
$I(k,k')=2\pi \int_0^\pi d \cos \theta ~\tilde C_1(k,k',\cos \theta),$ and
$J(k,k')=2\pi \int_0^\pi \cos \theta ~d \cos \theta
~\tilde C_1(k,k',\cos \theta), $ 
where
\begin{equation}
\tilde F_i(\vec k,\vec k')=\int \left[f(r)V_\tau( r)f( r)\right]~
e^{i(\vec k-\vec k') \cdot \vec r}~d\vec r \quad \quad \mbox{and}\quad\quad
\tilde C_1(\vec k,\vec k')=\int (f^2(r)-1)~
e^{i(\vec k-\vec k') \cdot \vec r}~d\vec r.
\end{equation}
More explicitly, the first terms of the above angle integrals
read for $f(r)=1$ :
\begin{eqnarray*}
A_\sigma&=&g_\sigma^2\theta_\sigma\quad
B_\sigma=g_\sigma^2\theta_\sigma\quad
C_\sigma=-2g_\sigma^2\phi_\sigma\\
A_\omega&=&2g_\omega^2\theta_\omega\quad
B_\omega=-4g_\omega^2\theta_\omega\quad
C_\omega=-4g_\omega^2\phi_\omega\\
\theta_i(p,p')&=&\log\left(\frac{(k+k')^2+m_i^2}{(k-k')^2+m_i^2}\right)\quad
\phi_i(p,p')=\frac{k^2+{k'}^2+m_i2}{4kk'}\theta_i(p,p')-1\,.
\end{eqnarray*}
The baryon density and the scalar density are
\begin{equation}
\rho_B =\frac{2~\nu}{(2\pi)^3}
\int_0^{k_f} d{\vec k}=\frac{2~\nu ~k_f^3}{6\pi^2},\quad\quad
\rho_s = \frac{2~\nu}{(2\pi)^3}
\int_0^{k_f} \cos\chi(\vec k)~d{\vec k}.
\end{equation}

\begin{figure}
\includegraphics[width=10.cm]{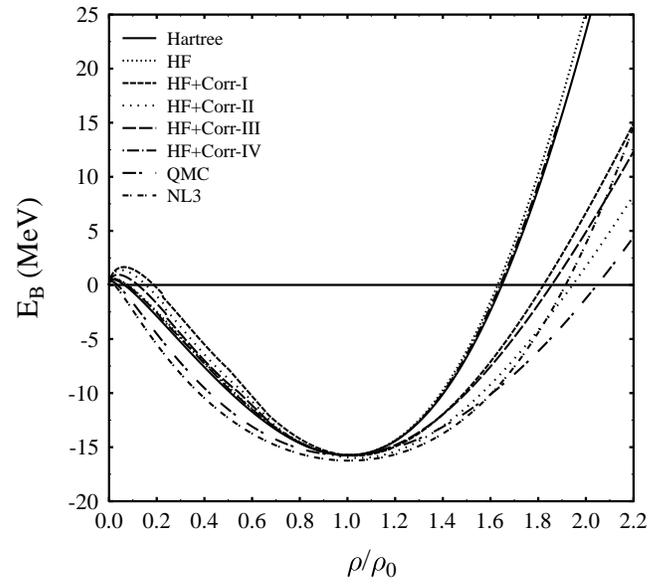}
\caption{EoS for different parametrizations as defined in the
table.} \label{eos}
\end{figure}

The parameters of the model have to be fixed. They are the
couplings $g_\sigma$, $g_\omega$, the meson masses, $m_\sigma$
and $m_\omega$ and
also three more parameters from the short range correlation
function, $\alpha$, $\beta$ and $\gamma$. The couplings 
are chosen so as to satisfy the ground state properties of
the nuclear matter and are given in table \ref{tab}. We choose 
$m_\sigma=550$ MeV and take $m_\omega=783$ MeV. 
The normalization condition (\ref{c0}) determines $\beta$. 
We fix $\alpha$ either by imposing the
condition $f(0)=0$, which appears to be a natural choice from our
experience with the non-relativistic case, or by minimizing
the energy. We choose $\gamma$ so that the correct healing
distance \cite{mosz} is reproduced. If $\alpha$ is chosen to be
1, we assume a density independent parameter $\gamma$ equal to 750
(HF+corr-IV). On the other hand, if we choose to
determine $\alpha$ variationally, we assume that the parameter
$\gamma$ increases linearly with the Fermi momentum. The last
choice is consistent with the idea that the healing distance
decreases as $k_F$ increases. Of course there are other possible
choices for these parameters. The parameters we have used are
tabulated in table \ref{tab} together with the compressibility $K$,
the relative effective mass $M^*/M$, the kinetic energy ${\cal T}/\rho_B-M$,
the direct and exchange parts of the potential energy (${\cal V}_d/\rho_B$
and ${\cal V}_e/\rho_B$ repectively) with correlation and the correlation 
contribution to the 
kinetic energy ${\cal T}^C/\rho_B$, all computed at the saturation point.
We next compute the range of the correlations, obviously related with the
healing distance, also included in table \ref{tab} and defined as \cite{bethe}:
\begin{equation}
R^2=\frac{{\int~r^2(f(r)-1)^2 ~d^3r}}{\int(f(r)-1)^2 ~d^3r}\,.
\end{equation}
In figure \ref{eos} we plot the binding energies as function of the
density for the Hartree, HF, quark-meson-coupling model (QMC) \cite{qmc}, 
a non-linear Walecka model NL3
\cite{nl3} and the 4 choices of the parameters in our calculation, as
given in table \ref{tab}. From this figure one can see that in
all cases that inclusion of correlations make the equation of
state (EOS) softer than Hartree or HF calculations. NL3 and QMC
also provide softer EOS around nuclear matter saturation density
but around two times saturation density, some of the EOS with
correlations are softer than NL3. Correlations always tend to
soften the EOS, except when $\alpha=1$ is fixed. In this case, the
EOS coincides with the curve obtained with the Hartree
calculation.

\begin{figure}
\includegraphics[width=10.cm]{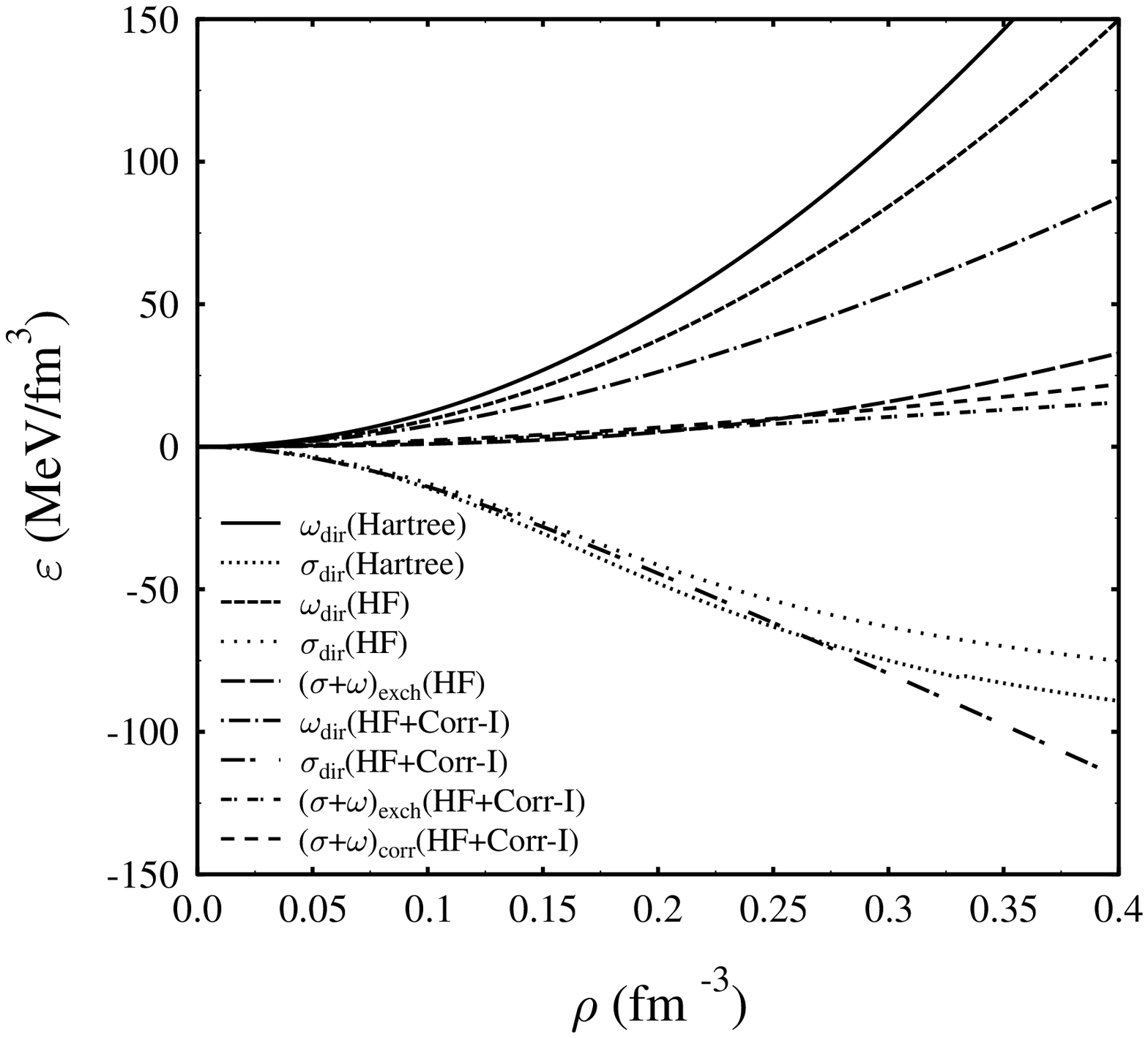}
\caption{Individual contributions for the energy density}
\label{indiv}
\end{figure}

In figure \ref{indiv} we plot the individual contributions to the energy
density in function of the density for the Hartree, HF and HF+correlations
(set III). Notice that the correlation contribution is of the same order as the
exchange contribution. Hence, it cannot be disregarded.
In fact if we compare the coupling constants $g_\sigma$ and $g_\omega$
obtained for the
different calculations we conclude that the introduction of the
correlations reduces the coupling constants. Correlation effects in the
Hartree and HF calculations are taken into account by a correct choice of
the coupling constants. However, the explicit introduction of correlations
has other effects such as softning the EOS.

In conclusion, we have included two-body correlations in the $\sigma-\omega$
relativistic model. We have seen that they correspond to an important amount
of the energy density and provide a softer EOS as compared with the HF
approximation. Furthermore, the effect of the correlations is similar to the
role played the non-linear 
terms in relativistic models or the 3-body forces in non-relativistic potential 
models. The inclusion of three-body correlations is currently under 
investigation.

This work was partially supported by CNPq (Brazil), CAPES (Brazil), GRICES
(Portugal) under project 100/03,  FEDER and FCT (Portugal) under the
projects POCTI/FP/FNU/50326/2003 and POCTI/FIS/451/94.

\begin{table}
\begin{ruledtabular}
\caption{Model parameters and ground state properties of nuclear
matter at saturation density are given. These results were obtained with
fixed: $M=939$ MeV $m_\sigma=550$ MeV, $m_\omega=783$
MeV at $k_{F0}=1.3$ fm$^{-1}$ with binding energy
$E_B=\varepsilon/\rho-M=-15.75$ MeV. We have used 4 choices of the
parameters for the correlation. HF+corr-I corresponds to
$600+200k_F/k_{F0}$, HF+corr-II to $700+200k_F/k_{F0}$,
HF+corr-III to $600+300k_F/k_{F0}$ and HF+corr-IV
corresponds to $\alpha=1$ and $\gamma$ fixed.} \label{tab}
\begin{tabular}{ccccccccccccc}
&$g_\sigma$ & $g_\omega$&$\alpha$& $\beta$& $\gamma$& R& K &
$M^*/M$& ${\cal T}/\rho_B -M$& ${\cal V}_d/\rho_B$&
${\cal V}_e/\rho_B$&${\cal T}^C/\rho_B$ \\
& & & & (MeV) & (MeV)&fm& (MeV)& & (MeV)&(MeV)&(MeV)&(MeV)\\
\hline
Hartree &11.079&13.806&  &  &  & &  540 & 0.540 & 8.11 &-23.86 &  & \\
HF &10.432&12.223&  &  &  & &  585 & 0.515 & 5.87 &-37.45 &15.83& \\
HF+corr-I&4.326&2.431& 13.474 &-1867.295 &800 &0.338 &523. & 0.568 &
14.96&-96.81 &26.77&39.33\\
HF+corr-II&4.359&2.4893& 13.655 &-2068.539 &900 &0.303&459. &0.608&
15.8&-83.01 &23.4&28.0\\
HF+corr-III&4.359&2.4893& 13.655 &-2068.539 &900 &0.303&448. &0.608&
15.8&-83.01 &23.4&28.0\\
HF+cor-IV&11.836&13.608&1.0 &-244.678&750  &0.318 &573. & 0.515&
7.16&-45.17 &17.94&4.31\\
\end{tabular}
\end{ruledtabular}
\end{table}

\begin{thebibliography}{99}
\bibitem{macgregor}M.H. MacGregor, A. R. Arndt, and R.M. Wright, Phys. Rev.
{\bf 182}, 1714 (1969) and references therein.
\bibitem{chpt} D.R. Entem and R. Machleidt, Phys. Lett {B 524}, 93 (2001).
\bibitem{bonn} R. Machleidt, Adv. Nucl. Phys. {\bf 19}, 189 (1989); R.
Machleidt, Phys. Rev. {\bf C 63}, 024001 (2001).
\bibitem{argonne} R. Wiringa, R. Smith, and T. Ainsworth, Phys. Rev {C 29}, 1207
(1984).
\bibitem{Coester} F. Coester, S. Cohen, B.D. Day, and C.M. Vincent, Phys. Rev.
{\bf C 1}, 769 (1970); R. Brockmann and R. Machleidt, Phys. Rev. {\bf C 42},
1965 (1990).
\bibitem{3-body} R.B. Wiringa, V. Fiks and A. Fabrocini, Phys. Rev. {\bf C 38},
1010 (1988); W. Zuo, A. Lejeune, U. Lombardo and J.-F. Mathiot,
Nucl. Phys. {\bf A 706}, 418 (2002).
\bibitem{teller} H.P. D\"urr and E. Teller, Phys. Rev. {\bf 101}, 494 (1956);
M. H. Johnson and E. Teller, Phys. Rev. {\bf 98}, 783 (1955); H.P. D\"urr,
Phys. Rev. {\bf 103}, 469 (1956).
\bibitem{walecka}B.D. Serot and J.D. Walecka, Adv. Nucl. Phys. {\bf 16}, 1
(1986); J.D. Walecka, Ann. of Phys. {\bf 83}, 491 (1974);
B.D. Serot, J.D. Walecka, Int. J. Mod. Phys. {\bf E6}, 
515 (1997).
\bibitem{horowitz} C.J. Horowitz and B.D. Serot, Nucl. Phys. {\bf A399},
529 (1983).
\bibitem{bouyssy}A. Bouyssy, J.-F. Mathiot and N.V. Giai and S. Marcos,
Phys. Rev. {\bf C 36}, 380 (1987).
\bibitem{villars}F. Villars, "Proceedings of the International School of
Physics, 'Enrico Fermi'-Course 23, (1961)."
Academic Press, New York, 1963; J.S. Bell, " Lectures on the Many-Body Problem, First Bergen International School of Physics." Benjamin,
New York, (1962); F. Coester and H. K\"ummel, Nucl. Phys. {\bf 17} 477 (1960).
\bibitem{jp}J. da Providencia and C.M. Shakin, Ann. Phys.(NY) {\bf 30},
95 (1964).
\bibitem{suzuki} K. Suziki, R. Okamoto and H. Kumagai, Phys. Rep. {\bf 242} 181
(1994).
\bibitem{feldmeier} H. Feldmeier, T. Neff, R. Roth, and J. Schnack, Nucl. Phys.
{\bf A 632}, 61 (1998); T. Neff and H. Feldmeier, Nucl. Phys. {\bf 713}
311 (2003).
\bibitem{mishra}A. Mishra, P.K. Panda, S. Schramm, J. Reinhardt and
W. Greiner, Phys. Rev. {\bf C 56}, 1380 (1997).
\bibitem{jp1}J. da Providencia and C. M. Shakin, Phys. Rev {\bf C 4}, 1560
(1971); C. M. Shakin, Phys. Rev {\bf C 4}, 684 (1971).
\bibitem{jastrow} R. Jastrow, Phys. Rev. {\bf 98}, 1479 (1955).
\bibitem{mosz}S.A. Moszkowski and B.L. Scott, Ann. Phys. (N.Y.), {\bf 11}, 65
(1960).
\bibitem{bethe} H. Bethe, Ann. Rev. Nucl. Sci. {\bf 21}, 93 (1971).
\bibitem{qmc} P. A. M. Guichon, Phys. Lett. {\bf B 200}, 235 (1988).
K. Saito and A.W. Thomas, Phys. Lett. B {\bf 327}, 9 (1994); P.K. Panda, 
A. Mishra, J.M. Eisenberg, W. Greiner, Phys. Rev. {\bf C 56}, 3134 (1997).
\bibitem{nl3} G. A. Lalazissis, J. K\"onig and P. Ring,
Phys. Rev. {\bf C 55}, 540 (1997).
\end{thebibliography}
\end{document}